\documentclass{article}
\usepackage{spconf}
\usepackage{amsmath,amssymb,mathtools}
\usepackage{graphicx}
\usepackage{booktabs,array}
\usepackage{subcaption}
\usepackage{multirow} 
\usepackage{hyperref}
\usepackage{placeins}

\usepackage{tikz}
\usetikzlibrary{arrows.meta,positioning,fit,calc}

\title{HyBeam: Hybrid Microphone--Beamforming Array-Agnostic Speech Enhancement for Wearables}
\name{Yuval Bar Ilan$^{1}$, Boaz Rafaely$^{1}$, Vladimir Tourbabin$^{2}$}
\address{
  $^{1}$School of Electrical and Computer Engineering, \\
  Ben-Gurion University of the Negev, Beer-Sheva 84105, Israel \\
  $^{2}$Reality Labs Research, Meta, Redmond, WA 98052, USA
}

\tikzset{
  >=Latex,
  font=\fontsize{9pt}{10.5pt}\selectfont,
  every node/.style={font=\fontsize{9pt}{10.5pt}\selectfont},
  block/.style={
    draw, rounded corners=2pt, align=center,
    inner sep=6pt, fill=gray!6,
    minimum width=4mm, minimum height=8mm
  },
  circ/.style={draw, circle, minimum size=7mm, inner sep=0pt},
  dashedsig/.style={densely dashed, line width=0.4pt},
  multisig/.style={->, line width=1.0pt}
}

\usepackage[protrusion=true,expansion=true]{microtype}

\renewcommand{\arraystretch}{0.95} 

\makeatletter
\renewcommand\section{\@startsection{section}{1}{\z@}%
  {6pt plus 2pt minus 2pt}
  {3.5pt plus 2pt minus 2pt}
  {\normalfont\large\bfseries\centering}}

\renewcommand\subsection{\@startsection{subsection}{2}{\z@}%
  {4pt plus 2pt minus 2pt}
  {2pt plus 2pt minus 2pt}
  {\normalfont\normalsize\bfseries}}

\renewcommand\subsubsection{\@startsection{subsubsection}{3}{\z@}%
  {3pt plus 2pt minus 2pt}
  {2pt plus 2pt minus 2pt}
  {\normalfont\normalsize\itshape}}
\makeatother

\begin{document}
\ninept
\maketitle

\begin{abstract}
Speech enhancement is a fundamental challenge in signal processing, particularly when robustness is required across diverse acoustic conditions and microphone setups. Deep learning methods have been successful for speech enhancement, but often assume fixed array geometries, limiting their use in mobile, embedded, and wearable devices. Existing array-agnostic approaches typically rely on either raw microphone signals or beamformer outputs, but both have drawbacks under changing geometries. We introduce HyBeam, a hybrid framework that uses raw microphone signals at low frequencies and beamformer signals at higher frequencies, exploiting their complementary strengths while remaining highly array-agnostic. Simulations across diverse rooms and wearable array configurations demonstrate that HyBeam consistently surpasses microphone-only and beamformer-only baselines in PESQ, STOI, and SI-SDR. A bandwise analysis shows that the hybrid approach leverages beamformer directivity at high frequencies and microphone cues at low frequencies, outperforming either method alone across all bands.

\end{abstract}
\begin{keywords}
array-agnostic, speech enhancement, beamforming, wearable arrays, hybrid models
\end{keywords}

\section{Introduction}
\label{sec:intro}
Speech enhancement (SE) aims to improve perceived quality and intelligibility in noisy, reverberant, and multi-speaker conditions, with applications such as teleconferencing, hearing aids, and voice interfaces \cite{hu2023se_survey}. Multichannel SE leverages spatial cues from microphone arrays, typically via classical beamformers (e.g., delay-and-sum or MVDR) combined with statistical post-filters, but these methods face limited performance in challenging acoustic scenes \cite{benesty2008mic, wang2024attn}. Recent advances in deep learning (DL) have substantially improved SE by modeling spectral, temporal, and spatial cues in a data-driven manner; however, most multichannel DL models remain tied to fixed array geometries (e.g., \cite{tolooshams2020caunet, wang2018comb, wang2018deepclust, wang2020dereverb}). To overcome this limitation, array-agnostic SE seeks to generalize across diverse array layouts and channel counts without retraining \cite{hsu2022,lee24}.

Several approaches have been proposed toward array-agnostic multichannel SE. One line of work processes each microphone stream independently with parameter sharing, followed by cross-stream aggregation, which simplifies deployment but underutilizes inter-channel spatial cues \cite{taherian2022}. To better capture spatial information, Transform--Average--Concatenate (TAC) modules were introduced \cite{luo2020tac,yoshioka2021vararray}, though they require multiple insertions across the network and significantly increase complexity. Another strategy is multi-geometry training, where models are exposed to diverse layouts during training \cite{zhang2021,taherian2022,pandey2021tadrn}, which reduces overfitting but demands large datasets and still does not guarantee generalization to unseen geometries.  Beamformer-based inputs have also been investigated in this context with the aim of reducing sensitivity to array geometry. For instance,  \cite{agadir2024} introduced array-geometry-agnostic processing based on beamforming for wearable head-mounted arrays, but performance was only investigated for automatic speech recognition rather than speech enhancement. Conversely, Yang et al. \cite{yang2023} studied beamforming-based speech enhancement, yet without addressing array-agnostic scenarios or robustness to microphone-position perturbations. Moreover, beamformers may be limited due to poor low-frequency directivity and high-frequency aliasing. Overall, existing work still leaves open key questions: (i) although both  microphone signals and beamformer outputs have been used as inputs to array-agnostic networks, no comprehensive analysis has clarified which is preferred under various conditions. (ii) In particular, for the relatively new form factor of microphone arrays embedded in smart glasses, the most effective strategy to achieve robustness against microphone-position perturbations has not yet been established.

To address these gaps, we present a comprehensive investigation leading to a hybrid design tailored for wearable arrays. Specifically, we combine raw microphone inputs at low frequencies with beamformer outputs at higher frequencies, leveraging their complementary strengths. This use of beamforming provides an initial spatial separation, which enables the use of compact networks suitable for edge devices, while keeping the framework strictly array-agnostic. The proposed hybrid approach is evaluated on both seen and unseen array geometries under microphone-position perturbations. The results demonstrate consistent improvements in perceptual quality (PESQ), intelligibility (STOI), and signal fidelity (SI-SDR) over baseline methods, showing that the proposed framework achieves superior array-agnostic robustness compared to models trained solely on raw microphones or on beamformers.

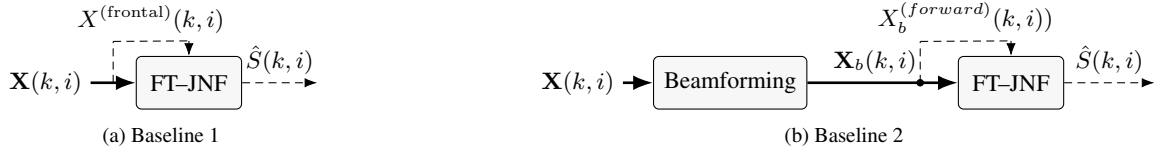
\begin{figure*}[t]
\centering

\begin{subfigure}[t]{0.49\textwidth}
\centering
\begin{tikzpicture}[
  node distance=7mm and 10mm,
  block/.style={draw, rounded corners=2pt, align=center,
                inner sep=4pt, fill=gray!6,
                minimum width=14mm, minimum height=7mm},
  circ/.style={draw, circle, minimum size=6mm, inner sep=0pt},
  dashedsig/.style={densely dashed, line width=0.4pt},
  multisig/.style={->, line width=1.0pt}
]
  \node[block] (ft) {FT--JNF};
  \node[left=6mm of ft.west] (xin) {$\mathbf{X}(k,i)$};
  \draw[multisig] (xin) -- (ft.west);

  \coordinate (lefttap) at ($(ft.west)+(-3mm,0)$);
  \coordinate (upPt)    at ($(lefttap)+(0,5.5mm)$);
  \coordinate (topPt)   at ($(ft.north)+(0,2mm)$);
  \draw[dashedsig] (lefttap) -- (upPt);
  \draw[dashedsig] (upPt) -- (topPt) node[midway, above] {$X^{(\mathrm{frontal})}(k,i)$};
  \draw[dashedsig,->] (topPt) -- (ft.north);

  \draw[dashedsig,->] (ft.east) -- ++(10mm,0) node[midway, above] {$\hat S(k,i)$};
\end{tikzpicture}
\caption{Baseline 1}
\label{fig:baseline1}
\end{subfigure}
\hfill
\begin{subfigure}[t]{0.49\textwidth}
\centering
\begin{tikzpicture}[
  node distance=7mm and 10mm,
  block/.style={draw, rounded corners=2pt, align=center,
                inner sep=4pt, fill=gray!6,
                minimum width=14mm, minimum height=7mm},
  circ/.style={draw, circle, minimum size=6mm, inner sep=0pt},
  dashedsig/.style={densely dashed, line width=0.4pt},
  multisig/.style={->, line width=1.0pt}
]
  \node[block] (bf) {Beamforming};
  \node[block, right=20mm of bf] (ft) {FT--JNF};
  \node[left=4mm of bf.west] (xin) {$\mathbf{X}(k,i)$};
  \draw[multisig] (xin) -- (bf.west);
  \draw[multisig] (bf.east) -- node[pos=0.45, above] {$\mathbf{X}_b(k,i)$} (ft.west);

  \path (bf.east) -- (ft.west) coordinate[pos=0.75] (tapxb);
  \node[fill, circle, inner sep=1pt] at (tapxb) {};
  \coordinate (upPt)  at ($(tapxb)+(0,5.5mm)$);
  \coordinate (topPt) at ($(ft.north)+(0,2mm)$);
  \draw[dashedsig] (tapxb) -- (upPt);
  \draw[dashedsig] (upPt) -- (topPt) node[midway , above] {$X_{b}^{(forward)}(k,i))$};
  \draw[dashedsig,->] (topPt) -- (ft.north);

  \draw[dashedsig,->] (ft.east) -- ++(12mm,0) node[midway, above] {$\hat S(k,i)$};
\end{tikzpicture}
\caption{Baseline 2}
\label{fig:baseline2}
\end{subfigure}

\caption{\textbf{Baseline models.} (a) \emph{Microphone input + microphone reference:} $\mathbf{X}_{\text{in}}(k,i)=\mathbf{X}(k,i)$; $X_{\text{ref}}(k,i)=X^{(\mathrm{frontal})}(k,i)$. \;
(b) \emph{Beamformer input + beamformer reference:} $\mathbf{X}_{\text{in}}(k,i)=\mathbf{X}_b(k,i)$; $X_{\text{ref}}(k,i)=X_{b}^{(forward)}(k,i))$.}
\label{fig:baselines}
\end{figure*}

\begin{figure*}[t]
\centering

\begin{subfigure}[t]{\textwidth}
\centering
\begin{tikzpicture}[node distance=10mm and 14mm, scale=0.95]
  \node[block] (bf) {Beamforming};
  \node[block, right=14mm of bf] (ft) {FT--JNF};
  \node[left=6mm of bf.west] (xin) {$\mathbf{X}(k,i)$};
  \draw[multisig] (xin) -- (bf.west);
  \draw[multisig] (bf.east) -- node[pos=0.4, above] {$\mathbf{X}_b(k,i)$} (ft.west);

  \path (xin) -- (bf.west) coordinate[pos=0.5] (tapmic);
  \node[fill, circle, inner sep=1pt] at (tapmic) {};
  \coordinate (upPt)  at ($(tapmic)+(0,6mm)$);
  \coordinate (topPt) at ($(ft.north)+(0,2mm)$);
  \draw[dashedsig] (tapmic) -- (upPt);
  \draw[dashedsig] (upPt) -- (topPt) node[midway, above] {$X^{(\mathrm{frontal})}(k,i)$};
  \draw[dashedsig,->] (topPt) -- (ft.north);

  \draw[dashedsig,->] (ft.east) -- ++(12mm,0) node[midway, above] {$\hat S(k,i)$};
\end{tikzpicture}
\caption{Hybrid 1}
\label{fig:hybrid1}
\end{subfigure}

\vspace{0.5em} 

\begin{subfigure}[t]{\textwidth}
\centering
\begin{tikzpicture}[
  node distance=10mm and 12mm,
  >=Latex,
  block/.style={draw, rounded corners=2pt, align=center,
                inner sep=6pt, fill=gray!6,
                minimum width=20mm, minimum height=9mm},
  circ/.style={draw, circle, minimum size=7mm, inner sep=0pt},
  every path/.style={rounded corners=1.2pt},
  scale=0.95
]

  \node (xin2) {$\mathbf{X}(k,i)$};
  \node[block, right=10mm of xin2] (bfbank) {Beamforming};
  \node[block, right=16mm of bfbank] (merge) {concatenation};
  \node[block, right=14mm of merge] (ft) {FT--JNF};

  \draw[multisig] (xin2.east) -- (bfbank.west);
  \draw[multisig] (bfbank.east) -- node[pos=0.45, above] {$\mathbf{X}_b(k,i)$} (merge.west);

  \coordinate (bypass) at ($(merge.west)+(-8mm,-6mm)$);
  \draw[multisig]
    (xin2.east) -- ++(3mm,0)
    |- (bypass)
    -- ($(merge.west)+(-8mm,-2mm)$)
    -- ($(merge.west)+(0,-2mm)$);

  \path (bfbank.east) -- (merge.west) coordinate[pos=0.80] (tapxb);
  \node[fill, circle, inner sep=1pt] at (tapxb) {};
  \draw[dashedsig] (tapxb) -- ++(0,8mm) coordinate (upxb);
  \draw[dashedsig,->] (upxb) -| (ft.north)
      node[midway, above, xshift=-18mm] {$X_{b}^{(forward)}(k,i))$};

  \draw[multisig] (merge.east) -- node[above] {$\mathbf{X}_{\text{hyb}}(k,i)$} (ft.west);
  \draw[dashedsig,->] (ft.east) -- ++(12mm,0) node[midway, above] {$\hat S(k,i)$};

\end{tikzpicture}
\caption{Hybrid 2}
\label{fig:hybrid2}
\end{subfigure}

\caption{\textbf{Proposed models.}
(a) \emph{Beamformer input + microphone reference (Hybrid~1):}
$\mathbf{X}_{\text{in}}(k,i)=\mathbf{X}_b(k,i)$; $X_{\text{ref}}(k,i)=X^{(\mathrm{frontal})}(k,i)$.
(b) \emph{Bandwise input + beamformer reference (Hybrid~2):}
$\mathbf{X}_{\text{in}}(k,i)=\mathbf{X}_{\text{hyb}}(k,i); 
X_{\text{ref}}(k,i)=X_{b}^{(forward)}(k,i)$.}
\label{fig:proposed}
\end{figure*}
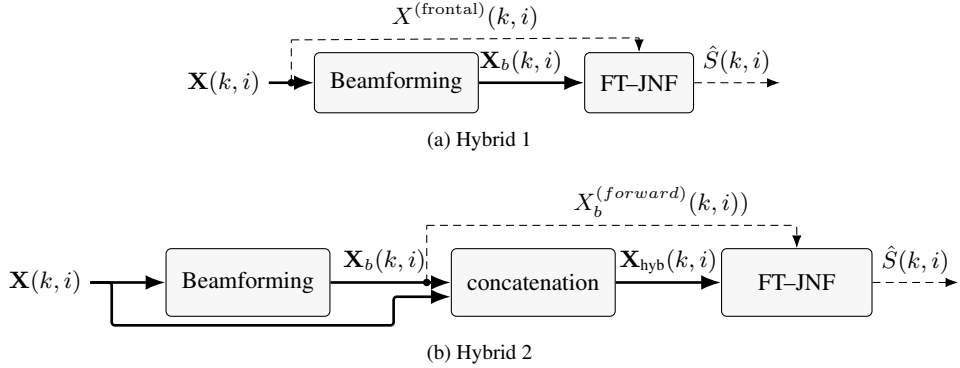

\section{Baseline Models}
\label{sec:baseline}

\subsection{Signal Model, Beamforming, and Masking Network}
We consider a clean target speech source signal $s(t)$, recorded by a wearable microphone array with $L$ channels in a multiple-speaker, noisy, and reverberant environment.  
The signal due to the target speaker at the $\ell$-th microphone, is denoted $y^{(\ell)}(t)$, and includes the effects of propagation delay and reverberation.
Transforming to the short-time Fourier transform (STFT) domain yields $Y^{(\ell)}(k,i)$, with frequency-bin index $k$ and time-frame index $i$.  

We denote by $V^{(\ell)}(k,i)$ the undesired component at microphone $\ell$, which subsumes both interfering speakers and additive microphone noise.  
The observed mixture at microphone $\ell$ is therefore
\begin{equation}
X^{(\ell)}(k,i) = Y^{(\ell)}(k,i) + V^{(\ell)}(k,i).
\end{equation}

Stacking all channels we obtain
\[
\mathbf{X}(k,i) = \mathbf{Y}(k,i) + \mathbf{V}(k,i) \in \mathbb{C}^L,
\]
where $\mathbf{Y}(k,i)=[Y^{(1)}(k,i),\ldots,Y^{(L)}(k,i)]^\top$ and $\mathbf{X}(k,i)$ similarly defined. 
For notational simplicity, we occasionally drop the explicit indices $(k,i)$ and denote by $\mathbf{Y} \in \mathbb{C}^{L \times F \times T}$ the full multichannel spectrogram, with $F$ frequency bins and $T$ time frames.  

In addition to the microphone signals, we also consider beamformed signals.  
Specifically, we employ a delay-and-sum (DAS) beamformer steered to a direction $d$ \cite{van1992beamforming}, whose output is
\begin{equation}
X_{b}^{(d)}(k,i) = \mathbf{w}_d^{H}(k)\, \mathbf{X}(k,i),
\end{equation}
where $\mathbf{w}_d(k)$ are the frequency-dependent DAS weights.  
Note that the explicit array geometry is never provided to the network. 
This keeps the learning component strictly array-agnostic.  
The collection of beamformer outputs across the selected directions is denoted 
\[
\mathbf{X}_b(k,i) = \{X_{b}^{(d)}(k,i)\}_{d \in \{\mathrm{1,...,D}\}},
\]
where $D$ corresponds to the number of beamformers.

The goal of speech enhancement (SE) is to recover the clean target waveform $s(t)$. As backbone we adopt the FT-JNF network~\cite{tesch2023deepfilters}, which estimates a complex ideal ratio mask (cIRM) $M(k,i)$.  
The network receives as input $\mathbf{X}_{\text{in}}(k,i)$, a set of multichannel spectrograms, and produces the mask.  
The enhanced signal is obtained by applying this mask to a chosen reference channel $X_{\text{ref}}(k,i)$:
\begin{equation}
\hat S(k,i) = M(k,i)\, X_{\text{ref}}(k,i).
\end{equation}
The time-domain enhanced signal $\hat s(t)$ is then reconstructed by applying the inverse STFT to $\hat S(k,i)$.  
The network is trained to minimize the scale-invariant SDR (SI-SDR) loss between the enhanced waveform $\hat s(t)$ and the clean target waveform $s(t)$.

\subsection{Baseline Model Parameterization}
\label{sec:params}
The baseline models described below, are based on the original FT-JNF network, but modified
by two design choices (cf. Fig.~\ref{fig:baselines}:

\begin{enumerate}
    \item \textbf{Network input} $\mathbf{X}_{\text{in}}(k,i)$: either the set of microphone signals $\mathbf{X}(k,i)$ or the set of beamformer outputs $\mathbf{X}_b(k,i)$.
    \item \textbf{Reference signal} $X_{\text{ref}}(k,i)$: either the frontal microphone $X^{(\text{frontal})}(k,i)$ with respect to the desired speaker position, or the forward-looking beamformer $X_{b}^{(forward)}(k,i)$.
\end{enumerate}

\subsection{Baseline 1: Microphones}
Baseline~1 is illustrated in Fig.~\ref{fig:baseline1}.  
The input is the microphone signals and the reference is the frontal microphone, i.e.  
$\mathbf{X}_{\text{in}}(k,i) = \mathbf{X}(k,i), \quad 
X_{\text{ref}}(k,i) = X^{(\text{frontal})}(k,i)$.

\subsection{Baseline 2: Beamforming}
Baseline~2 is illustrated in Fig.~\ref{fig:baseline2}.  
The input is the beamformer outputs and the reference is the forward-looking beam, i.e.  
$\mathbf{X}_{\text{in}}(k,i) = \{X_{b}^{(d)}(k,i)\}_{d \in \{\mathrm{1,...,D}\}}, \quad 
X_{\text{ref}}(k,i) = X_{b}^{(forward)}(k,i)$. We used four steering directions for the beamformers (front, back, left, right), with weights computed from the array microphone positions (at both train and test time).

\section{Proposed Hybrid Models}
\label{sec:hybrid}
The baseline models use either microphone inputs or beamformer outputs exclusively - representations that have been previously studied as possible inputs to masking networks. 
However, it is not clear, particularly for the wearable array form factor studied here, 
which representation is preferable and under what conditions. 
We therefore propose hybrid configurations (cf. Fig.~\ref{fig:proposed}) 
that combine both types of signals in different combinations, complementing the baselines, as detailed below. The hybrid models are illustrated in Fig.~\ref{fig:proposed}.

\subsection{Proposed Model Parametrization}
The proposed hybrid models are distinguished by different design choices based on the following parameters:
\begin{enumerate}
\item \textbf{Bandwise hybrid input:}  
Performance analysis presented in Sec.~\ref{sec:exp2_res} of the baseline methods reveals that Baseline~1 performs better at low frequencies, while Baseline~2 at high frequencies. 
This motivated a hybrid model whose input uses microphone signals at low frequencies and beamformer signals at high frequencies. Formally, the hybrid input is
\begin{equation}
\mathbf{X}_{\text{hyb}}(k,i) =
\begin{cases}
\mathbf{X}(k,i), & k < k_c, \\[6pt]
\mathbf{X}_b(k,i), & k \geq k_c,
\end{cases}
\end{equation}
where $\mathbf{X}(k,i)$ denotes the multichannel microphone STFTs, 
$\mathbf{X}_b(k,i)$ the set of beamformer outputs, 
and $k_c$ is the frequency bin corresponding to the cutoff frequency $f_c$, 
selected based on validation performance ($f_c = 1500$\,Hz).

\item \textbf{Hybrid Microphones -- Beamforming (channel-wise):}  
While the baseline methods employed either microphones or beamforming signals, this hybrid design choice mixes both in a single network. 
In particular, we combine a reference microphone signal with beamformer outputs as network input. 
The reference channel involves a trade-off: applying the mask to a beamformer output provides spatial selectivity but may add artifacts, while applying it to a raw microphone signal avoids distortions but lacks spatial filtering.

\end{enumerate}

The hybrid models are as detailed in Fig.~\ref{fig:proposed}.

\subsection{Hybrid 1: Beamforming}
Hybrid~1 is illustrated in Fig.~\ref{fig:hybrid1}.  
The input is the beamformer outputs and the reference is the frontal microphone, i.e.  
$\mathbf{X}_{\text{in}}(k,i) = \mathbf{X}_{b}(k,i), \quad 
X_{\text{ref}}(k,i) = X^{(\text{frontal})}(k,i)$.

\subsection{Hybrid 2: Bandwise + Beamforming}
Hybrid~2 is illustrated in Fig.~\ref{fig:hybrid2}.  
The input is the bandwise hybrid (mics low, beamformers high) and the reference is the forward-looking beam, i.e.  
$\mathbf{X}_{\text{in}}(k,i) = \mathbf{X}_{\text{hyb}}(k,i), \quad 
X_{\text{ref}}(k,i) = X_{b}^{(forward)}(k,i)$.

\subsection{Hybrid 3: Bandwise + Microphones}
Hybrid~3 follows the bandwise design of Hybrid~2 (see Fig.~\ref{fig:hybrid2}); it differs only in the reference channel, the mask is applied to the frontal microphone rather than to the forward-looking beam. 
Formally,
$\mathbf{X}_{\text{in}}(k,i) = \mathbf{X}_{\text{hyb}}(k,i), \quad
X_{\text{ref}}(k,i) = X^{(\text{frontal})}(k,i)$.

\section{Experimental setup and methodology}
\label{sec:setup}
\begin{figure}[t]
  \centering
  \includegraphics[width=4cm,trim=0 0cm 0 0cm,clip]{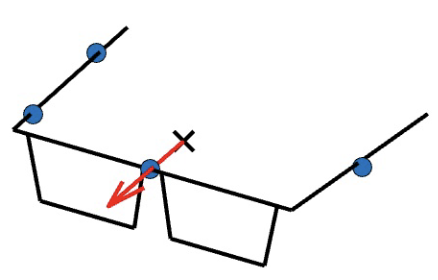}
  \vspace{-2mm}
    \caption{Array~0 (nominal), with microphone marked by the blue dots. 
    Microphones positions (in mm) (x,y,z) =: $(-29,\,82,\,-5)$, $(30,\,-1,\,-1)$, $(11,\,-77,\,-2)$, $(-60,\,-83,\,-5)$. 
    The forward-looking axis (positive x-axis) is shown relative to the glasses’ center. 
    The array center is marked by a black ``X", with position (x,y,z)=(0,0,0).}
  \label{fig:array0_nominal_3d}
  \vspace{0mm}
\end{figure}

\begin{figure}[t]
  \centering
  \hspace{0.05\linewidth} 
  \begin{subfigure}{0.23\linewidth}
    \centering
    \includegraphics[width=\linewidth]{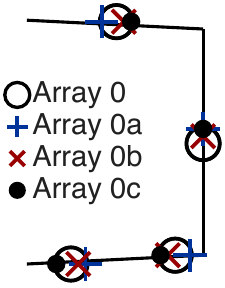}
    \subcaption{}
  \end{subfigure}
  \hspace{0.1\linewidth} 
  \begin{subfigure}{0.23\linewidth}
    \centering
    \includegraphics[width=\linewidth]{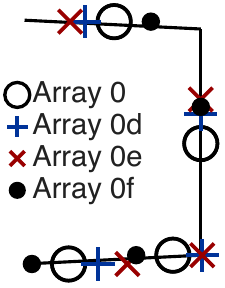}
    \subcaption{}
  \end{subfigure}
  \hspace{0.05\linewidth} 

  \vspace{0.3cm}

  \hspace{0.05\linewidth} 
  \begin{subfigure}{0.23\linewidth}
    \centering
    \includegraphics[width=\linewidth]{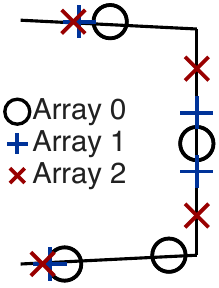}
    \subcaption{}
  \end{subfigure}
  \hspace{0.1\linewidth} 
  \begin{subfigure}{0.23\linewidth}
    \centering
    \includegraphics[width=\linewidth]{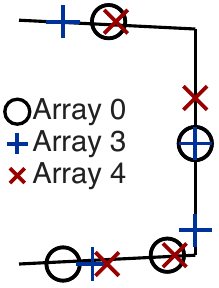}
    \subcaption{}
  \end{subfigure}
  \hspace{0.05\linewidth} 

  \caption{Arrays~0a–0f and 1–4 (top view). 
    (a) Small perturbations (5–10 mm)
    (b) Large perturbations (20–40 mm)
    (c) and (d) Different geometries.}
  \label{fig:arrays_grouped}
\end{figure}

\subsection{Room Simulation}

We generate simulated room impulse responses (RIRs) using Pyroomacoustics~\cite{pyroomacoustics}, which implements the image source method (ISM)~\cite{allen1979image}. 
For each example, the room dimensions are sampled independently with room length $L\!\sim\!\mathcal{U}(2.5,5.0)$\,m, width $W\!\sim\!\mathcal{U}(3.0,9.0)$\,m, and 
height $H\!\sim\!\mathcal{U}(2.2,3.5)$\,m, and the reverberation time is drawn from $T_{60}\!\sim\!\mathcal{U}(0.2,0.5)$\,s.

The wearable array is composed of  four microphones, and is assumed to be mounted on the frame of glasses, as illustrated in Fig.~\ref{fig:array0_nominal_3d}.
For each example, the array center, marked "x" in the figure,  is placed at least $1$\,m from the walls, at position $(x,y,z)$ drawn from $x\!\sim\!\mathcal{U}(1,L{-}1)$ , $y\sim\!\mathcal{U}(1,W{-}1)$ and a fixed height of $z{=}1.5$\,m. 
The glasses frame’s rotation about the vertical axis is drawn uniformly from $[0,2\pi)$. 

Once the RIRs are defined, we place speech sources in the simulated room. 
Each example contains one frontal target talker and five interferers, all modeled as point sources. 
The target is located at $0^\circ$ azimuth relative to the forward-looking axis of the glasses (see Fig.~\ref{fig:array0_nominal_3d}), at a distance $r_s\!\sim\!\mathcal{U}(0.3,1.0)$\,m in the horizontal plane (height $1.5$\,m). 
The azimuth range $20^\circ$--$340^\circ$ is divided into five equal sectors, and one interferer is placed at a random azimuth within each sector. 
Their distances follow $r_i \!\sim\! \mathcal{U}(1,8)$\,m.
Their heights are sampled from a normal distribution with mean $1.6$\,m and standard deviation about of 0.28 m . 
The signals are sampled at 16\,kHz taken from the WSJ0 corpus~\cite{garofolo1993wsj0}. 

Microphone signals are then obtained by convolving the source signals with the corresponding RIRs at each microphone position. 
We apply the short-time Fourier transform (STFT) with a Hann window of $N_{\text{fft}}{=}512$ samples and $50\%$ overlap (256-sample hop). 
Finally, sensor noise is added to yield an input $\mathrm{SNR }$ of $ 30$\,dB.

\subsection{Array Configurations}
We consider 11 arrays in total.  
Array~0 serves as the unperturbed reference geometry (Fig.~\ref{fig:array0_nominal_3d}).  

Two sets of perturbations are defined relative to Array~0:
\begin{itemize}
  \item \textbf{Small perturbations (5–10 mm):} Arrays~0a, 0b, 0c.  
  \item \textbf{Large perturbations (20–40 mm):} Arrays~0d, 0e, 0f.  
\end{itemize}

In addition, Arrays~1–4 represent substantially different geometries, not derived directly from Array~0. Illustrations are provided in Fig.~\ref{fig:arrays_grouped}, where each perturbed array is shown alongside the nominal reference.

\subsection{Experiments methodology}
We define two main experiments, each with its own training, validation, and test setup. 
Experiment~1 uses only Array~0 for training and validation, and evaluates on Array~0 together with its perturbed variants (0a–0f), thereby isolating robustness to perturbations when training is restricted to the nominal geometry. 
Experiment~2 uses training and validation sets constructed from a diverse subset of eight arrays, spanning both perturbation groups and alternative geometries. 
The test set, in contrast, includes the full collection of arrays (Arrays~0--4 and 0a--0f), which divides into \emph{seen} arrays (present in training/validation) and \emph{unseen} arrays (excluded from training, and included arrays 0c, 1, 4). 
This setup enables a clear evaluation of array-agnostic generalization. 
For this experiment, we also report bandwise SI-SDR results to analyze the frequency-dependent contribution of microphone and beamformer cues.

\subsection{Training Details}
All models are trained using the Adam optimizer with learning rate $1 \times 10^{-3}$ for up to 100 epochs. 
The objective function is the scale-invariant SDR (SI-SDR) loss~\cite{leroux2019sdr}, and the best checkpoint is selected based on validation SI-SDR.  

\subsection{Evaluation}
Performance is assessed on both seen and unseen geometries to quantify array-agnostic robustness. 
Metrics include scale-invariant SDR (SI-SDR)~\cite{leroux2019sdr}, perceptual evaluation of speech quality (PESQ)~\cite{itu2001pesq}, and short-time objective intelligibility (STOI)~\cite{taal2010stoi}, each computed against clean speech.

\section{Results and Discussion}
\label{sec:results}

\subsection{Experiment 1: Nominal array design}
Baselines are trained on the reference geometry (Array~0) and evaluated on three groups: 
the reference (Array~0), small perturbations (Arrays~0a–0c), and large perturbations (Arrays~0d–0f) .The results are presented in Table 1, and show that the microphone-based baseline (Baseline~1) attains the best STOI, PESQ, and SI-SDR measures for the reference array and for the small perturbations arrays, 
indicating that raw microphone cues remain reliable when geometry deviations are mild. 
With larger geometry perturbations, the beamformer-based baseline (Baseline~2) becomes superior in all three metrics, 
suggesting that beamformer inputs are less sensitive to microphone-position shifts in this regime.

\begin{table}[t]
\centering
\caption{Experiment~1 (Nominal array design): Reference vs.\ perturbation groups (best in \textbf{bold}). 
Groups: Ref = Array~0; Small = \{0a–0c\}; Large = \{0d–0f\}. 
SI = SI-SDR.}
\label{tab:ds1_groups}

\begingroup
\setlength{\tabcolsep}{2pt}        
\renewcommand{\arraystretch}{0.95} 
\fontsize{9pt}{10.5pt}\selectfont  

\begin{tabular}{@{}l *{3}{ccc} @{}}
\toprule
\multirow{2}{*}{Model}
  & \multicolumn{3}{c}{\textbf{Ref}}
  & \multicolumn{3}{c}{\textbf{Small}}
  & \multicolumn{3}{c}{\textbf{Large}} \\
& {\fontsize{8.4pt}{10pt}\selectfont STOI} 
& {\fontsize{8.4pt}{10pt}\selectfont PESQ} 
& {\fontsize{8.4pt}{10pt}\selectfont SI}
& {\fontsize{8.4pt}{10pt}\selectfont STOI} 
& {\fontsize{8.4pt}{10pt}\selectfont PESQ} 
& {\fontsize{8.4pt}{10pt}\selectfont SI}
& {\fontsize{8.4pt}{10pt}\selectfont STOI} 
& {\fontsize{8.4pt}{10pt}\selectfont PESQ} 
& {\fontsize{8.4pt}{10pt}\selectfont SI} \\

\midrule
NOISY       & 0.57 & 1.12 & -11.2 & 0.57 & 1.12 & -10.9 & 0.57 & 1.11 & -11.1 \\
Baseline 1  & \textbf{0.82} & \textbf{1.56} & \textbf{1.1}
            & \textbf{0.82} & \textbf{1.52} & \textbf{1.1}
            & 0.60 & 1.19 & -8.2 \\
Baseline 2  & 0.80 & 1.50 & 0.2
            & 0.79 & 1.48 & 0.1
            & \textbf{0.70} & \textbf{1.27} & \textbf{-4.5} \\
\bottomrule
\end{tabular}

\endgroup
\end{table}

Overall, microphone-only inputs are preferable near the nominal geometry, whereas beamformer-only inputs are more robust under larger perturbations. 
These complementary trends motivate the hybrid models introduced in Sec.~\ref{sec:hybrid}.

\subsection{Experiment 2: Multiple array design}
\label{sec:exp2_res}

In this experiment, training is conducted using a diverse subset of arrays sampled from both perturbation groups and the substantially different geometries. 
Evaluation covers the full set of arrays (Arrays~0–4 and 0a–0f), enabling assessment of generalization to unseen geometries (arrays 0c, 1, 4 in this case).

\begin{table}[t]
\centering
\caption{Experiment~2: Bandwise SI\,-\,SDR values in dB (averaged across all arrays). 
Hybrid~2 and Hybrid~3 use a cutoff frequency of $f_c=1.5$\,kHz. Freq. bands for each column are presented in kHz.}
\label{tab:bandwise_ds2}

\begingroup
\setlength{\tabcolsep}{5pt}        
\renewcommand{\arraystretch}{0.95} 
\fontsize{9pt}{10.5pt}\selectfont  

\begin{tabular}{lccccc}
\toprule
Model & \multicolumn{5}{c}{\textbf{Frequency (kHz)}} \\ 
\cmidrule(lr){2-6}
       & 0--0.5 & 0.5--1 & 1--2 & 2--4 & 4--8 \\
\midrule
NOISY       & -9.2 & -9.3 & -12.5 & -16.9 & -18.0 \\
Baseline 1  &  3.7 &  3.5 &  -1.8 & -13.1 & -37.3 \\
Baseline 2  &  2.7 &  3.2 &  -1.6 & -10.2 & -25.1 \\
Hybrid 1    &  2.6 &  2.8 &  -2.4 & -12.1 & -23.1 \\
Hybrid 2    &  \textbf{4.0} & \textbf{3.8} & \textbf{-1.5} & \textbf{-10.3} & \textbf{-19.1} \\
Hybrid 3    &  \textbf{4.0} & \textbf{3.8} &  -2.2 & -12.2 & -29.1 \\
\bottomrule
\end{tabular}

\endgroup
\end{table}

Table~\ref{tab:bandwise_ds2} reports bandwise SI-SDR results averaged across all arrays. 
The bandwise SI-SDR is computed by isolating the relevant frequency band in the STFT domain for both the estimated signal and the clean reference, transforming each band back to the time domain, and then applying the SI-SDR measure. 
At \textbf{low frequencies ($\leq 1$ kHz)}, the microphone-based baseline (Baseline~1) achieves better performance than the beamformer-based baseline (Baseline~2), highlighting the advantage of direct microphone inputs in this regime. 
At \textbf{mid-to-high frequencies (1--4 kHz)}, the trend reverses: Baseline~2 outperforms Baseline~1, indicating that beamformer inputs provide better spatial separation in this band.
At the \textbf{highest band (4--8 kHz)}, none of the models improve upon the noisy input, though Baseline~2 still surpasses Baseline~1 in line with the mid-frequency results. 
Since speech energy in this range is very low, the outcomes have minimal influence on the overall SI-SDR. 
The hybrid models (Hybrid~2 and Hybrid~3) exploit the strengths of both input types across all frequency bands, achieving superior performance to both baselines at low, mid, and high frequencies alike.

\begin{table}[t]
\centering
\caption{Experiment~2: Averages by array groups. 
Seen arrays vs.\ unseen arrays. Hybrid~2 and Hybrid~3 use a cutoff frequency of $f_c=1.5$\,kHz.}
\label{tab:ds2_overall_avg}

\begingroup
\setlength{\tabcolsep}{4.5pt}        
\renewcommand{\arraystretch}{0.95} 
\fontsize{9pt}{10.5pt}\selectfont  

\begin{tabular}{l*{2}{ccc}}
\toprule
\multirow{2}{*}{Model}
  & \multicolumn{3}{c}{\textbf{Seen}}
  & \multicolumn{3}{c}{\textbf{Unseen}} \\
& STOI & PESQ & SI-SDR & STOI & PESQ & SI-SDR \\
\midrule
NOISY       & 0.57 & 1.12 & -11.1 & 0.58 & 1.12 & -11.0 \\
Baseline 1  & 0.80 & 1.52 &   0.7 & 0.80 & 1.54 &   0.7 \\
Baseline 2  & 0.81 & 1.56 &   0.5 & 0.81 & 1.55 &   0.2 \\
Hybrid 1    & 0.80 & 1.55 &   0.3 & 0.80 & 1.54 &   0.0 \\
Hybrid 2    & \textbf{0.82} & \textbf{1.63} & \textbf{1.0} & \textbf{0.82} & \textbf{1.65} & \textbf{1.1} \\
Hybrid 3    & 0.81 & 1.59 &   0.9 & 0.81 & 1.59 &   1.0 \\
\bottomrule
\end{tabular}

\endgroup
\end{table}

Table~\ref{tab:ds2_overall_avg} reports average results separately for arrays seen during training and for unseen arrays. 
Performance on the unseen arrays shows no degradation relative to the seen arrays across STOI, PESQ, and SI-SDR, 
indicating that multi-geometry training preserves array-agnostic behavior. 
Overall, hybrid models outperform the baselines in PESQ and SI-SDR for both seen and unseen array groups. 
Hybrid~2 achieves the best STOI, PESQ, and SI-SDR across both groups, outperforming all baselines and the other hybrid models.
Importantly, the performance of Hybrid~2 across all arrays is comparable to or better than that of Baseline~1 trained in Experiment~1 on the nominal reference array, demonstrating that combining the hybrid design with training on diverse arrays enables array-agnostic generalization without loss in nominal performance.

\section{Conclusion }
\label{sec:Conclusion}
We presented \textbf{HyBeam}, a hybrid microphone--beamforming framework for array-agnostic speech enhancement on wearables. By combining raw microphones at low frequencies with beamformer outputs at high frequencies, it exploits complementary spatial cues without exposing array geometry. Experiments showed that hybrid models outperform microphone- or beamformer-only baselines.
Future work could focus on exploring optimal beamformers for the task and improving performance in the high-frequency range (4--8 kHz).

\FloatBarrier

\bibliographystyle{IEEEbib}
\bibliography{refs}

\end{document}